                                                                  \documentclass[12pt,preprint]{aastex}
\shorttitle{The formation of two active-region filaments}
\shortauthors{Yan et al.}

\begin{document}

\title{Successive X-class flares and coronal mass ejections driven by shearing motion and sunspot rotation in active region NOAA 12673}
\author{X. L. Yan\altaffilmark{1,2,3}, J. C. Wang\altaffilmark{1,4}, G. M. Pan\altaffilmark{5}, D. F. Kong\altaffilmark{1}, Z. K. Xue\altaffilmark{1}, L. H. Yang\altaffilmark{1}, Q. L. Li\altaffilmark{1,4}, X. S. Feng\altaffilmark{6} }

\altaffiltext{1}{Yunnan Observatories, Chinese Academy of Sciences, Kunming 650011, China. yanxl@ynao.ac.cn}
\altaffiltext{2}{Sate Key Laboratory of Space Weather, Chinese Academy of Sciences, Beijing 100190, China.}
\altaffiltext{3}{Center for Astronomical Mega-Science, Chinese Academy of Sciences, 20A Datun Road, Chaoyang District, Beijing, 100012, China.}
\altaffiltext{4}{University of Chinese Academy of Sciences, Yuquan Road, Shijingshan Block Beijing 100049, China.}
\altaffiltext{5} {College of Mathematics Physics and Information Engineering, Jiaxing University, Jiaxing 314001, China.}
\altaffiltext{6} {SIGMA Weather Group, State Key Laboratory for Space Weather, National Space Science Center,
Chinese Academy of Sciences, Beijing 100190, China.}

\begin{abstract}
We present a clear case study on the occurrence of two successive X-class flares including a decade-class flare (X9.3) and two coronal mass ejections (CMEs) triggered by shearing motion and sunspot rotation in active region NOAA 12673 on 2017 September 6. A shearing motion between the main sunspots with opposite polarities started on September 5 and even lasted after the second X-class flare on September 6. Moreover, the main sunspot with negative polarity rotated around its umbral center and another main sunspot with positive polarity also exhibited a slow rotation. The sunspot with negative polarity at the northwest of active region also began to rotate counter-clockwise before the onset of the first X-class flare, which is related to the formation of the second S-shaped structure. The successive formation and eruption of two S-shaped structures were closely related to the counter-clockwise rotation of three sunspots. The existence of a flux rope is found prior to the onset of two flares by using non-linear force free field extrapolation based on the vector magnetograms observed by SDO/HMI. The first flux rope corresponds to the first S-shaped structures mentioned above. The second S-shaped structure was formed after the eruption of the first flux rope. These results suggest that shearing motion and sunspot rotation play an important role in the buildup of the free energy and the formation of flux ropes in the corona which produces solar flares and CMEs.

\end{abstract}

\keywords{Sun: sunspots - Sun: activity - Sun: photosphere - Sun: flares - Sun: coronal mass ejections (CMEs)}

\section{Introduction}
Sunspot rotation was discovered  about one century ago (Evershed 1910; Maltby 1964). Using high spatial and temporal resolution of recent ground-based and satellite-borne telescopes, many sunspots were found to rotate around the center of its umbra or another sunspot in one solar cycle (Nightingale et al. 2001; Brown et al. 2003; Yan et al. 2008a; Zheng et al. 2016).

Sunspot rotation is very effective to twist and shear magnetic field lines, which is believed to be one of the sources that non-potential energy comes from. As is well known,  solar flares and CMEs are caused by a sudden release of free magnetic energy that was previously stored in the magnetic fields (Forbes 2000; Priest 2002; Chen 2011;Shen et al. 2011; Deng et al. 2013a; Lin et al. 2015; Wang \& Liu 2015; Tian et al. 2015; Zhang et al. 2015; Li, Innes \& Ning 2016; Xue et al. 2016; Kors{\'o}s \& Erd{\'e}lyi 2016; Dhara et al. 2017;  Kors{\'o}s, Ruderman \& Erdelyi, R.\ 2018). The free magnetic energy can be transformed into the kinetic energy of particles, radiation, and heating during solar eruptions.  Wang et al. (1996) studied the evolution of vector magnetic fields with solar flares and found that the occurrence of solar flares is closely associated with sheared magnetic fields and great enhancement of vertical currents. 

Recently, R{\'e}gnier \& Canfield (2006) found that the storage of magnetic energy caused by a slow rotation of a sunspot in NOAA 8210 is enough to power for a C-class flare. Furthermore, interaction between a fast rotating sunspot and ephemeral regions was found to trigger the X-class (Zhang et al. 2007). Yan et al. (2008b) first classified the rotating sunspots into six types in solar cycle 23 and investigated solar flare productivity of each type rotating sunspot. They found that rotation direction of the sunspots opposite to the differential rotation in active regions can produce higher strong (X-class) flares. Using continuum intensity images and vector magnetograms from the Spectropolarimeter (SP) of Hinode, Yan et al. (2009) found that an counterclockwise rotation of the sunspot with positive polarity in active region NOAA 10930 occurred before an X3.4 flare. Jiang et al. (2012) studied the sunspot rotation associated with an X-class flare in AR 11158 on 2011 February 15 in solar cycle 23. Through calculating the energy in the corona produced by the sunspot rotation, Kazachenko et al. (2009) and Li \& Liu (2015) suggested that sunspot rotation is an effective mechanism for building up magnetic energy in the corona. More and more observation confirmed that sunspot rotation can trigger solar eruptions (Yan \& Qu 2007; Zhang, Liu, \& Zhang 2008; Li, \& Zhang 2009; Vemareddy, Ambastha, \& Maurya 2012; Deng et al. 2013b; Dhara et al. 2014; Chen et al. 2014; Gopasyuk 2015; Li, \& Liu 2015; Suryanarayana et al. 2015; Vemareddy, Cheng, \& Ravindra 2016; Wang et al. 2016; Zhang et al. 2017). The formation and eruption of S-shaped active-region filaments or flux ropes was also found to be associated with sunspot rotation (Yan et al. 2012, 2015; Yang et al. 2015; Ruan et al. 2015; James et al. 2017; Li et al. 2017). Furthermore, the S-shaped magnetic structures are often associated with solar flares and CMEs (Canfield et al. 2007). When a twisted flux rope emerges from below the photosphere to coronal heights, the footpoints of the flux rope in the photosphere also exhibit rotation motion (Amari et al. 1996; Kumar et al. 2013; Yan et al. 2017). Interestingly, Moon et al. (2002) found that the sign of the impulsive helicity variation is opposite to that of the initial smooth helicity variation during the flares. Liu et al. (2016) reported the sudden flare-induced rotation of a sunspot due to the surface Lorentz-force change caused by the back reaction of coronal magnetic restructuring using the unprecedented spatiotemporal resolution of the 1.6 m New Solar Telescope. The abrupt reversal of rotation in the sunspot also was observed during an X1.6 flare (Bi et al. 2016).

Except for the observational evidence that sunspot rotation is closely associated with solar eruptions, many simulations were performed to confirm that sunspot rotation can twist the magnetic field line and forms the twisted or sigmoid magnetic structure (T{\"o}r{\"o}k et al. 2003; Magara 2006; Fan 2009). T$\ddot{o}$r$\ddot{o}$k et al. (2013) found that twisting driven by sunspot rotation can lead to the expansion of the overlying field that can reduce magnetic tension and lead to the filament eruption. Sturrock et al. (2015) performed a 3D MHD numerical experiment of the emergence of a toroidal flux tube from the solar interior through the photosphere and into the solar corona. In their simulation, they found that sunspot rotation within the two polarity sources during the magnetic emergence can leave the interior portion of the field untwisted and twist up the atmospheric portion of the field. Sturrock \& Hood (2016) found that the faster a sunspot rotates, the more magnetic helicity and energy were transformed to the atmosphere. Sunspot rotation can lead to the formation of the twisted magnetic structure (Yan et al. 2015). The footpoints of a twisted flux rope  also exhibit rotation motion during its emergence (Min \& Chae 2009; MacTaggart \& Hood 2009; Hood, Archontis \& MacTaggart 2012).

In this paper,  we present the evolution of  sunspots in active region NOAA 12673 associated with two X-class flares and two CMEs in detail. The details of the observations are presented in section 2. The results are shown in section 3. The conclusion and discussion are given in section 4.

\section{Observations}

The data used in the paper are mainly from Atmospheric Imaging Assembly (AIA; Lemen et al. 2012) and the Helioseismic and Magnetic Imager (HMI; Schou et al. 2012) on board the Solar Dynamics Observatory (SDO). The AIA can provide multiple, simultaneous high-resolution full-disk images of the transition region and the corona. The spatial and temporal resolutions of AIA are 1.$^\prime$$^\prime$5 and 12 s, respectively. The 304 \AA, 171 \AA, 131 \AA\ images observed by SDO/AIA are employed to show the  eruption process of two X-class flares.  The HMI can provide full disk continuum intensity image and line-of-sight magnetograms at 45 s cadence with a precision of 10 G. The continuum intensity images and line-of-sight magnetograms obtained by SDO/HMI are used to show the evolution of sunspots in active region NOAA 12673 from 2017 September 5 to September 6. 

The vector magnetograms observed by HMI on board the SDO (Schou et al. 2012; Bobra et al. 2014; Centeno et al. 2014) were employed to show the evolution of longitudinal and transverse magnetic fields. These magnetograms from Space Weather HMI Active Region Patch (SHARP) series have a pixel scale of about 0.$^\prime$$^\prime$5 and a cadence of 12 minutes. The Very Fast Inversion of the Stokes Vector algorithm (Borrero et al. 2011) was used to derive the vector magnetic fields. The minimum energy method (Metcalf 1994; Metcalf et al. 2006; Leka et al. 2009) is used to resolve the 180 degree azimuthal ambiguity. The images are remapped using Lambert (cylindrical equal area) projection centered on the midpoint of the active region, which is tracked at the Carrington rotation rate (Sun 2013).

The TiO images used in this letter are observed by the New Vacuum Solar Telescope (NVST)(Liu et al. 2014). The TiO images have a pixel size of 0.$^\prime$$^\prime$04 and a cadence of 12 s. These data are calibrated from Level 0 to Level 1 with dark current subtracted and flat field corrected, and then speckle masking method was used to reconstruct the calibrated images from Level 1 to Level 1+ by  (Weigelt 1977; Lohmann et al. 1983; Xiang et al. 2016).  Moreover, TiO images observed by the NVST are used to show the sunspot structure in the photosphere before the first X-class flare. 

All the SDO data are calibrated to Level 1.5 by using the standard procedure in SSW, and rotated differentially to a reference time (at 06:15:00UT on 2017 September 6). Then, we co-aligned the SDO and NVST images using the cross-correlation method (Feng et al. 2015; Yang et al. 2015; Xiang et al. 2016).

\section{Result}
Active-region NOAA 12673 appeared at the east solar limb on 2017 August 31 with an $\alpha$ field configuration of the sunspot group. It seems very stable from 2017 August 31 to September 2. However, it developed rapidly on September 3 and the main sunspot with positive polarity expanded much larger than before and many small sunspots with negative polarity emerged from the north and the east of the main sunspot (Yang et al. 2017; Sun \& Norton 2017). The sunspot with negative polarity at the north of the main sunspot moved from the east to west and became larger. The active region developed into a $\beta$$\gamma$$\delta$ magnetic field configuration of the sunspot group on 2017 September 6.

Two X-class flares occurred on 2017 September 6 (see the GOES X-ray profile in Fig. 1(a)). The black line indicates the profile of 1-8 \AA\ and the blue line indicates the profile of 0.5-4 \AA\ from 00:00 UT to 24:00 UT on 2017 September 6. The information of solar flares can be extracted from GOES flare catalogue as follows: The first X-class (X2.2) began at  08:57 UT, peaked at  09:10 UT, and ended at 09:17 UT. The second X-class flare (X9.3) began at 11:53 UT, peaked at 12:02 UT, and ended at 12:10 UT. Up to now, the latter flare is the biggest flare in solar cycle 24. Figure. 1(b) shows the whole active region observed in the continuum intensity image superimposed by the contour of line-of-sight magnetogram. The line-of-sight magnetogram of the active region observed by HMI was shown in Figure. 1(c). The contour levels of the magnetic fields in Fig. 1(b) are $\pm$200 G, $\pm$1000 G, and $\pm$1800 G. The red and blue contours indicate the positive and negative magnetic polarities, respectively. During the evolution of the active region, three sunspots labeled as S1, S2 and S3 exhibited an obvious rotation motion. The blue box in Fig. 1(c) outlines the field of view of Figs. 2, 3, 5, 6 and 9, and the yellow box outlines the field of view of Figs. 7 and 8. Sunspots S1 and S2 approached together and formed a $\delta$ sunspot on 2017 September 6 (see high resolution observation of TiO images in Figs 1(d) and 1(e)). The structure of sunspot S2 was very complex. There were two light bridges in sunspot S2. Sunspot S1 squeezed tightly the left part of sunspot S2. 

In order to present the evolution of the active region, we calculate the velocity fields by using DAVE method and the vector magnetic fields. Figure 2 shows the evolution of sunspots S1 and S2 in the continuum intensity images, the line-of-sight magnetograms, and the velocity fields superimposed on the continuum intensity images from the left column to the right one before the first X-class flare (see animation movie 1 that includes the evolution of the active region during the period of the two flares). From the evolution of these sunspots, sunspot S1 moved toward the north of sunspot S2. Next, sunspot S1 began to squeeze the northeast part of sunspot S2. The yellow dotted lines outline the change of the bright structure between sunspots S1 and S2 (see Figs. 1(a), 1(d), and 1(g)). This bright structure exhibited an S-shaped structure. Except for the movement of sunspot S1 toward the north of sunspot S2, the north part of sunspot S1 also rotated around the center of its umbra. Sunspot S2 also exhibited a slow rotation motion. The method to measure the rotation is described below. The penumbra between sunspots S1 and S2 changed into hook-shaped structure from the line-shaped structure. Sunspot S1 developed into a tadpole-shaped structure. Sunspots S1 and S2 were rotating counter-clockwise. In order to calculate the rotation angle of sunspot S1, we draw a circle covering the north part of sunspot S1 and trace the evolution of the north part of the S-shaped structure (see the dotted line in Figs. 2(a), (d), and (g)) with time. The red lines mark the position of the head of the S-shaped structure that rotated around the center of the north part of sunspot S1. At the first four hours (from 00:03:34 UT - 04:00:34 UT), the sunspot S1 exhibit a slow rotation and the rotation angle is about 10 degrees. The average rotation speed reached 2.5 degrees per hour. However, the sunspot S1 experienced a rapid rotation near the onset of the first flare (from 04:00:34 UT - 09:08:04 UT). The rotation angle of sunspot S1 reached about 50 degrees before the first X-class flare. The average rotation speed reached 10 degrees per hour. The evolution of line-of-sight magnetograms is shown in Figs. 1(b), 1(e), and 1(H). Sunspot S1 intruded gradually into sunspot S2 at the north part of sunspot S2. The green arrows in Figs. 2(c), 2(f), and 2(i) show the directions of the transverse flow fields. The shearing motion between sunspots S1 and S2 can be seen from 03:48 UT to 04:36 UT. Afterwards, the sunspot S1 began to rotate. Sunspot rotation can be seen clearly in the velocity map (see Fig. 2(i)). The velocity field of sunspots S1 and S2 exhibited a vortex shape. The positions of the enhancement of the intensity in the continuum intensity image were marked by two blue arrows at the maximum of the X9.3 flare in Fig. 2(g). The X2.2 flare was a white light flare (Neidig, Wiborg \& Gilliam 1993; Ding et al. 1999; Song et al. 2018) due to the emission in the continuum images during the flare.

Based on the vector magnetograms observed by SDO/HMI, we extrapolated the three-dimensional structure of the active region. The change of the magnetic structure can be seen in Fig. 3. At the beginning of the shearing motion between the two sunspots, the magnetic structure exhibited a sheared arcade configuration (see Fig. 3a).  After about eight hours, the sheared arcades evolved into a flux rope (see Fig. 3b). The average twist of the selected magnetic field lines is about 1.1 $\pi$. Figs. 3(c) and 3(d) show the magnetic structures from left side of Figs. 3(a) and 3(b).  

Figure 4 shows the eruptive process of the X2.2 flare in 304 \AA, 171 \AA, and 131 \AA\ images from the left column to the right one. All the images were overlaid by the contours of the line of sight magnetograms. The green and blue contours indicate the positive and negative magnetic polarities. The contour levels of the magnetic fields are $\pm$350 G, $\pm$1000 G, and $\pm$1800 G. Before the X2.2 flare occurred, the structure of this active region in the 304 \AA\ and 171 \AA\ observation has no obvious change. The sheared magnetic loops between sunspots S1 and S2 can be seen from a sequence of 131 \AA\ observation. There was a filament between sunspots S1 and S2 along the polarity inversion line (see animation movie 2 from 08:56 UT to 09:02 UT). The eruption of the S-shaped filament between sunspots S1 and S2 produced the first X-class flare. The location of the S-shaped filament corresponds to that of the S-shaped flux rope extrapolated from nonlinear force free field. Therefore, the filament is constrained in the twisted flux rope. The sigmoid structure in the corona corresponds to the twisted flux rope. At the maximum of the flare, there was an S-shaped bright ribbon that just located between sunspots S1 and S2 (see Fig. 3(g), 3(h), and 3(i)). The chromosphere and corona evolution of the X2.2 flare can be seen from animation movie 2.

After the first X-class flare occurred, sunspot S1 experienced a rapid rotation. The yellow dotted lines outline the change of the S-shaped structure between sunspots S1 and S2 (see Figs. 5(a), (d), and (g)). The S-shaped structure mentioned here referred to the structure seen in the photosphere, which is different from the sigmoid structure observed in soft X-ray observation. The difference of the filament, sigmoid structure, and highly twisted flux tube was investigated by R{\'e}gnier \& Amari (2004). From the continuum intensity images, sunspot S1 rotated about 10 degrees from 09:00:34 UT to 10:30:34UT (see the change of the red line in Figs. 5(a) and 5(d)). The average rotation speed is about 6.6 degrees per hour. The sunspot S1 rotated about 15 degrees during about one and half an hour before the onset of the second X-class flare. The average rotation speed is about 10 degrees per hour. The evolution of line of sight magntograms was shown in Figs. 5(b), 5(e), and 5(h). Up to the occurrence of the second X-class flare, the north part of sunspot S1 almost broke the north part of sunspot S2 into two parts. The green arrows indicate the direction of the flow field. The shearing motion and sunspot rotation can be also judged from the velocity map. Especially, the flow fields of sunspots S1 and S2 exhibited an obvious vortex shape. The positions of the enhancement of the intensity in the continuum intensity image were marked by two blue arrows at the maximum of the X9.3 flare in Fig. 5(g). The X9.3 flare was also a white light flare. The evolution of sunspots S1 and S2 can be seen from animation movie 1.

The evolution of vector magnetograms of sunspots S1 and S2 is shown in Fig. 6. The red arrows show the transverse magnetic fields of negative polarity and the blue arrows show the transverse magnetic fields of positive polarity. The magnetic field lines between sunspots S1 and S2 have a strong shear angle along polarity inversion line. The transverse fields of the sunspots S1 and S2 exhibit a strong departure from potential field. Especially, the transverse fields at the north part of the sunspot S1 are around its center of the umbra.

Figure 7 shows the evolution of sunspot S3 in the continuum intensity image, line-of-sight magnetogram, and flow field map. At 05:06 UT, light bridge began to form in the umbra of sunspot S3 with negative polarity. After the light bright appeared, the light bridge changed its shape into Y-shaped structure at about 07:48 UT. The sunspot S3 began to rotate just before the occurrence of the first X-class flare. In the following, the light bridge began to rotate around the center of sunspot S3. The rotation of the sunspot S3 can be deduced from the change of the light bridge (see the red dotted lines in Figs. 7(a), 7(d), 7(g)). The rotation angle of sunspot S3 reached about 50 degrees. The average speed is 16.6 degrees per hour. The evolution of line-of-sight magntograms was presented in Figs. 7(b), 7(e), and 7(h). The flow fields also exhibited an obvious vortex shape as those of sunspots S1 and S2 (see Figs. 7(i)). The evolution of sunspot S3 can be seen from animation movie 1. The evolution of vector magnetograms of sunspot S3 is shown in Fig. 8. The red arrows show the transverse magnetic fields of negative polarity and the blue arrows show the transverse magnetic fields of positive polarity. The transverse fields of the sunspots S1 and S2 exhibit a vortex shape. 

After the first X-class flare, the north part of the sunspot S1 still rotated around its center of the north part. The sunspot S1 separated the north part of sunspot S2 into two parts (see Figs .6d, 6e and 6f). The three-dimensional structure of the active region was extrapolated by using NLFFF model based on the vector magnetograms. The average twist of the selected magnetic field lines (the pink curved lines) formed  at 11:12 UT along the polarity inversion line is about 1.4 $\pi$ (see Fig. 9). Figure 9 shows the twisted magnetic structure between sunspots S1 and S2. It implies that the second flux rope was formed after the first flare between sunspots S1 and S2. The configuration of the second flux rope is very similar to the first one (see Figs. 3b and 3d). Figs. 9(a) and 9(b) show the structure of the second flux rope seen from top view and from left side. 

Two and a half hours after the occurrence of the first X-class flare, this active region produced the second X-class flare that is the biggest flare in solar cycle 24 up to now. Figure 10 shows the eruptive process of the X9.3 flare in 304\AA, 171\AA, and 131\AA\ images from the left column to the right one. All the images were overlaid by the contours of the line of sight magnetograms. The green and blue contours indicate the positive and negative magnetic polarities. The contour levels of the magnetic fields are $\pm$350 G, $\pm$1000 G, and $\pm$1800 G. Before the occurrence of the second X-class flare, an inverse S-shaped structure can be seen in 131 \AA\ observation (see the red dotted line in Fig. 10(i)), which is corresponding to the second flux rope extrapolated from nonlinear force free field. There were three parts of magnetic loops marked by the red dotted lines in Fig. 10(c). The left two magnetic loops (a small inverse S-shaped loop between sunspots S1 and S2 and an arcade loop in the northwest of sunspot S1) first reconnected. The small inverse S-shaped structure corresponds to the twisted flux rope obtained by using NLFFF extrapolation. A newly inverse S-shaped magnetic loop formed after reconnection (see the left red dotted line in Fig. 10(f)). Next, the newly formed inverse S-shaped magnetic loop reconnected with another larger arcade loop that connected sunspot S2 and sunspot S3 as same as the event studied by Tian et al. (2017). Finally, a large inverse S-shaped magnetic structure formed (see the red dotted line in Fig. 10(i)). The inverse S-shaped structure can be only seen in 131 \AA\ observation. It implies that this magnetic structure may be a hot channel corresponding to a flux rope. The chromosphere and corona evolution of the X9.3 flare can be seen from animation movie 2.

Figure 11 displays the evolution of the magnetic flux, magnetic helicity injection,  and magnetic helicity accumulation of the whole active region from 00:00 UT to 15:00 UT on 2017 September 6. The two groups of the vertical dashed red lines indicate the starting and ending time of the two X-class flares. The blue and black lines indicate the negative and positive magnetic flux of the whole active region in Fig. 11(a). The negative magnetic flux decreased and the positive polarity magnetic flux increase from 00:00 UT to 06:21 UT on September 6. There is a data gap from 06:21 UT to 07:41 UT due to the fuzziness of the magnetograms. The negative and positive magnetic flux began to decrease at 07:41 UT at the same time. After the first X-class flare, both of the negative and positive magnetic flux increase slowly and then decrease. The negative magnetic flux increases rapidly after the X9.3 flare. There is a common characteristic that both of the negative and positive magnetic flux decrease before the two X-class flares. The phenomenon implies that magnetic cancellation occurred before occurrence of the flares.

The buildup of the non-potential energy can be deduced from magnetic helicity injection and accumulation (Liu, Zhang \& Zhang 2008). In order to investigate the change of the magnetic helicity associated with the flares, the line-of-sight HMI magnetograms were used to calculate the magnetic helicity injection rate and helicity accumulation of the whole active region. The magnetic helicity injection rate can be calculated by using the following equation:

\begin{equation}
\frac{dH}{dt}=-2\oint(A_p \cdot u)B_ndS,
\end{equation}

where $A_p$ is the vector potential of the potential field and $u$ is the photospheric transverse velocity field derived by using DAVE method (Schuck 2006). $B_n$ is the normal component of the magnetic field to the photospheric surface S. The method is similar to that of  Chae (2001) and Liu \& Schuck (2013). 

As the rotation motion of sunspots S1, S2, and S3 was counterclockwise, the magnetic helicity of the whole active region can be induced to be negative. According to the change of the helicity rate and helicity accumulation in Fig. 10(b), the continuous magnetic helicity was injected into the upper atmosphere from the photosphere before the occurrence of the two flares. It is obvious that sunspot rotation can transform the non-potentiality energy from the photosphere to the upper atmosphere.  Note that the changes in helicity during the flare can be induced by an instrumental effect. From the evolution of the line-of-sight magnetograms, the measurements of the magnetograms are affected during the flares from 09:00 UT to 09:36 UT and from 11:54 UT to 12:18 UT. But most of helicity injection rate is negative except for two short time periods during the flares, which resulted from the instrument. Therefore, the accumulation helicity is negative for the whole active region.

There were two CMEs associated with the two X-class flares observed by LASCO C2 on 2017 September 6. Figs. 11(a) and 11(b) show the first CME associated with the first X-class flare. Figs. 11(c) and 11(d) show the second CME associated with the second X-class flare. The LASCO C2 images were superimposed by the full disk 304 \AA\ images. The bright region in the south hemisphere in 304 \AA\ images was the location of active region NOAA 12673. The first CME was corresponding to the eruption of the small inverse S-shaped magnetic structure between sunspots S1 and S2. The second CME was corresponding to the eruption of the large inverse S-shaped magnetic structure connecting sunspots S2 and S3. Therefore, the second CME is much larger than the previous one. To identify the relationship between the flares and CMEs, we use the linear fitting and quadratic curve fitting to the height of CMEs with time. The time difference of the onset of the first (second) CME and the onset of the first (the second) flare  is within two hours. During this period, there is no other flares occurring in this active region. Therefore, the first and second flares are associated with the first and second CMEs.

\section{Conclusion and discussion}
Active region NOAA 12673 produced two successive X-class flares from 08:57 UT to 12:10 UT on 2017 September 6. Up to now, the X9.3 flare was the biggest flare in solar cycle 24. Before the occurrence of the two flares, Two main sunspots with opposite polarities exhibited a clear counter-clockwise rotation and there was also a shearing motion between them. An S-shaped structure formed between two sunspots and its eruption produced the first X-class flare. Except for the rotation of two sunspots, another sunspot with negative polarity in the northwest of two main sunspots also had counter-clockwise rotation. The second X-class flare (9.3) was caused by the eruption of a large inverse S-shaped sigmoid structure that formed due to the magnetic reconnection between the small S-shaped magnetic structure and the other part of the magnetic field lines that was connecting to the rotating sunspot in the northwest of the two main rotating sunspots. The helicity of this active region was negative, which is consistent with the counter-clockwise rotation of the sunspots. Moreover, we found that the existence of a flux rope between the main sunspots is found before the onset of each flare. It is also found that the sunspot rotated faster than before when the active region approached the production of the flares.

Three sunspots in this active region exhibited counter-clockwise rotation, which is opposite to the direction of the differential rotation. Furthermore, the accumulation magnetic helicity of this active region was negative as a whole. The positive helicity injection rate during the occurrence of the flares is due to the effect from the instrument. Yan et al. (2008b) found that active regions hosting rotating sunspots that the rotation direction of these sunspots opposite to the differential rotation of the Sun have higher X-class flare production. Active region NOAA 12763 was located in the south hemisphere of the Sun. The event studied in this paper confirmed the results obtained by Yan et al. (2008b). More and more observational evidences reveal that if the active regions have rotating sunspot, the production rate of solar flares is much higher than the ones that have not rotating sunspots (Yan et al. 2009; Jiang et al. 2012). The first X-class flare is closely associated with the eruption of a small inverse S-shaped magnetic structure between sunspots S1 and S2. As two legs of this magnetic structure rooted in the two sunspots with opposite magnetic polarities that had counterclockwise rotation. The twist can be injected into the small inverse S-shape magnetic structure by sunspot rotation and its eruption produced the first X-class flare. After the occurrence of the first X-class flare, sunspots S1 and S2 were still rotating. The slow rising of the magnetic loop connecting two rotating sunspots reconnected with other part magnetic loop and formed a new S-shaped structure. The final formed inverse S-shaped magnetic structure was connecting sunspots S2 and S3. Without doubt, reconnected magnetic structures have twist in them due to the rotation of these sunspots. Based on the vector magnetograms observed by SDO/HMI, we find that there is a flux rope between sunspots S1 and S2 prior to the onset of each flare by using non-linear force free field extrapolation method. The existence of a flux rope is often found prior to solar eruptions (Kliem et al. 2013; Liu et al. 2014; Liu et al. 2016; Shen et al. 2018). We assume that the formation of the flux ropes is closely related to sunspot rotation.  Before the appearance of the first flux rope, the shearing motion was found along the PIL between the two sunspots and the sheared arcade loops were formed due to the shearing motion. After the sunspot rotation appeared, the first flux rope was found prior to the first X-class flare. It is obvious that the twist of the flux rope came from the sunspot rotation. The process of the second flare is very similar to that of the first flare.

Consequently, sunspot rotation may play a very important role in accumulation of magnetic energy and helicity. Even there are some simulations performed about the sunspot rotation to obtain the process of solar eruption, it is also poorly understood the nature of  what causes sunspot rotation and the relationship between sunspot rotation and solar eruptions. Especially, why do sunspots rotate faster approaching the onset of the eruptions? In future, the development of helioseismology may reveal the truth of sunspot rotation in the solar interior.

\acknowledgments
The authors thank the referee for her/his constructive suggestions and comments that helped to improve this paper. We would like to thank the NVST, SDO/AIA, and SDO/HMI teams for the high-cadence data support. This work is sponsored by the National Science Foundation of China (NSFC) under the grant numbers 11373066, 11603071, 11503080, 11633008, 11533008, by the Youth Innovation Promotion Association CAS (No.2011056), by Project Supported by the Specialized Research Fund for State Key Laboratories and by the grant associated with project of the Group for Innovation of Yunnan Province.

\begin{figure}
\epsscale{.65}
\plotone{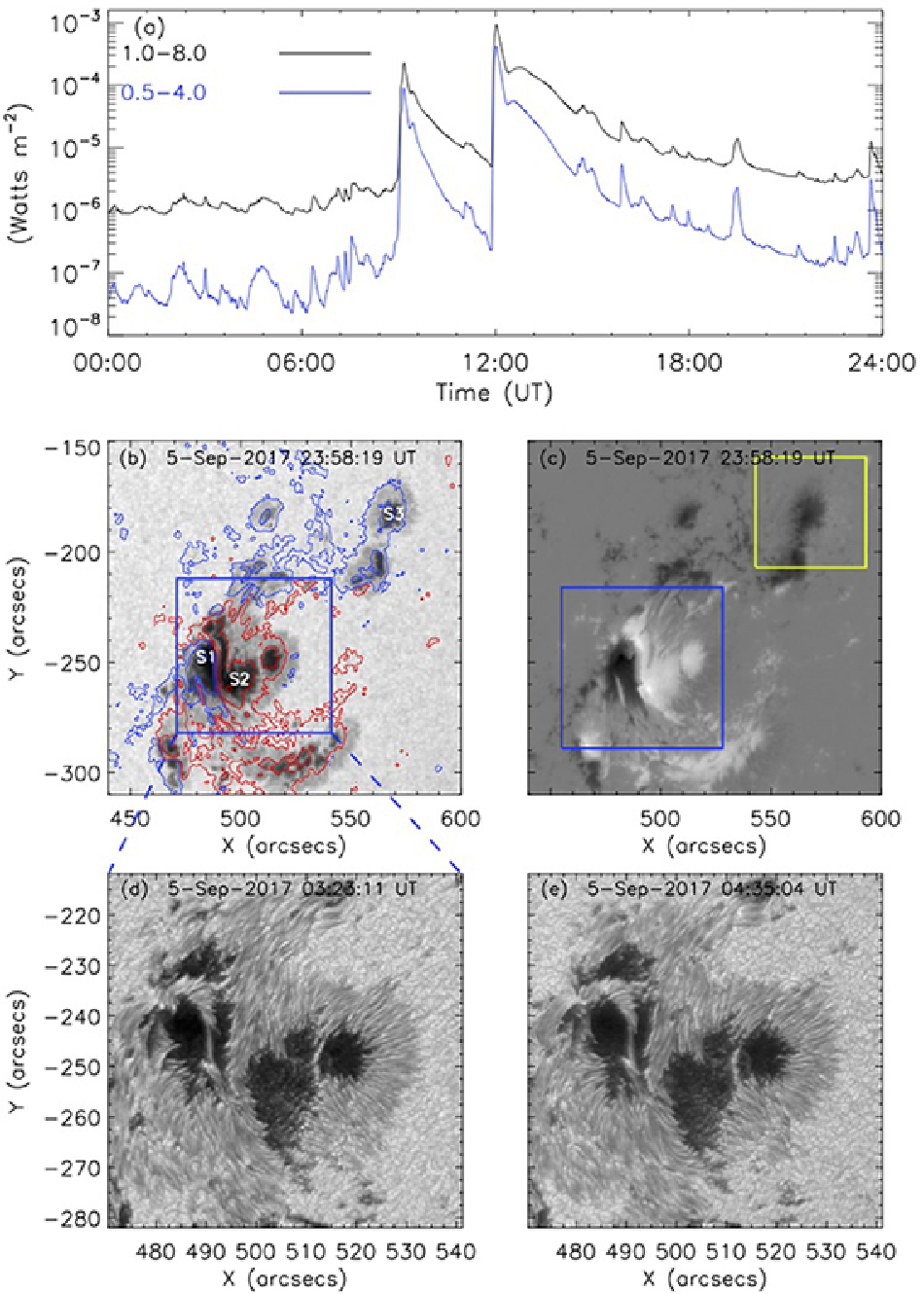}
\caption{GOES X-ray profile, continuum intensity image,  line of sight magnetogram, and TiO images. (a): GOES X-ray profile. The black line indicates the profile of 1-8 \AA\ and the blue line indicates the profile of 0.5-4 \AA\ from 00:00 UT to 24:00 UT on 2017 September 6. (b): The continuum intensity image superimposed by the contours of the line of sight magnetogram. The contour levels of the magnetic fields in the right panel are $\pm$200 G, $\pm$1000 G, and $\pm$1800 G in Fig. 1(b). (c): Line of sight magetogram. The red and blue contours indicate the positive and negative magnetic polarities.  The blue box in Fig. 1(c) outlines the field of view of Fig. 2 and  Fig.4 and the yellow box outlines the field of view of Figs. 5. (d) and (e): Hight resolution TiO images observed by the NVST. \label{fig1}}
\end{figure}

\begin{figure}
\epsscale{.90}
\plotone{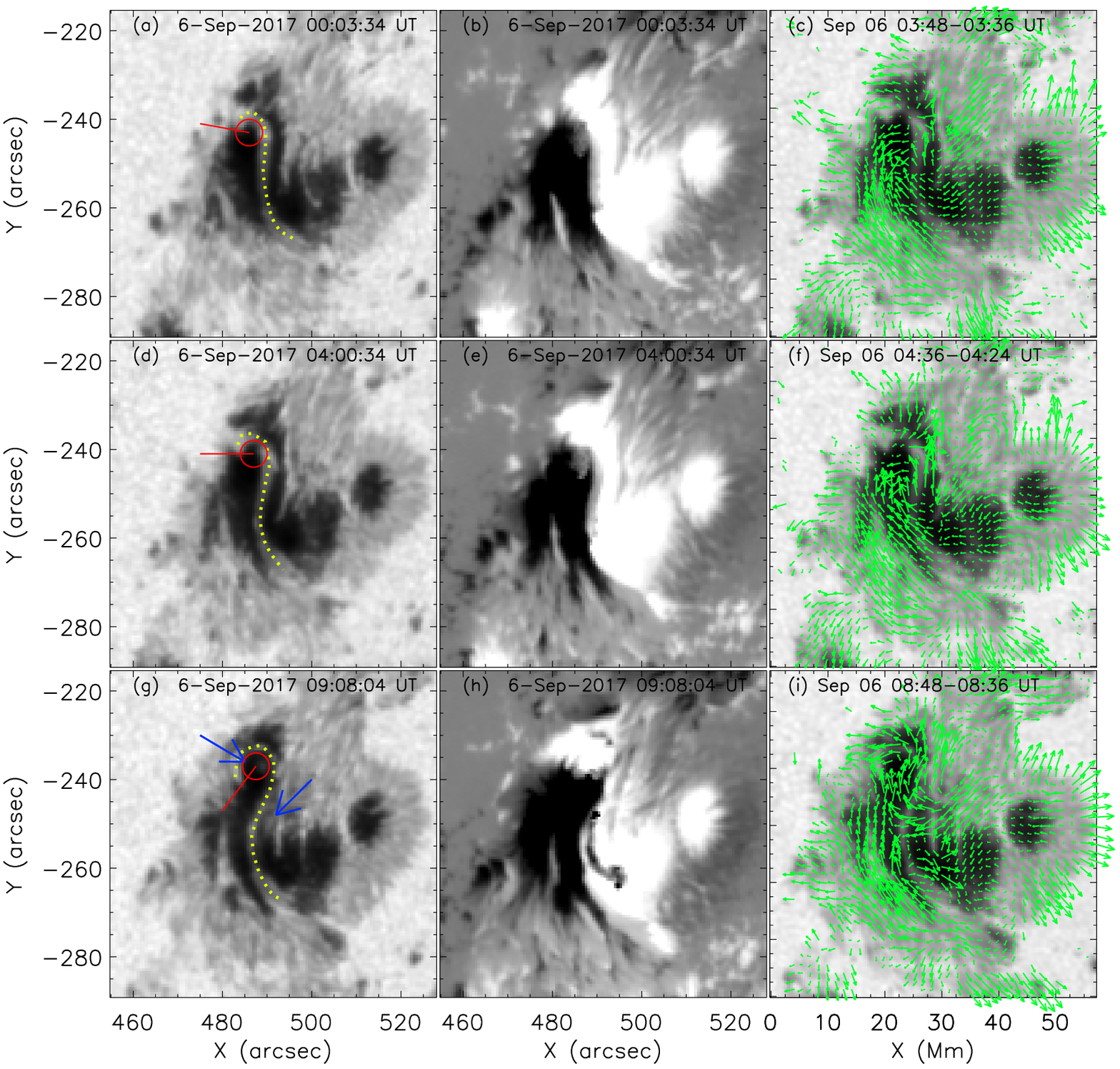}
\caption{Evolution of sunspots S1 and S2 in the continuum intensity images (the left column), the line of sight magnetograms (the middle column), and the velocity fields superimposed on the continuum intensity images (the right column). The yellow dotted lines indicate the change of the shape of the bright structure between sunspots S1 and S2. The green arrows in Figs. 2(c), 2(f), and 2(i) show the directions of the transverse flow fields. The positions of the enhancement of the intensity in the continuum intensity image were marked by two blue arrows at the maximum of the X2.2 flare in Fig. 2(g). The red circle covers the north part of sunspot S1 and is used to trace the rotation of the north part of the S-shaped structure with time. The red lines mark the position of the head of the S-shaped structure that rotated around the center of the north part of sunspot S1. \label{fig1}}
\end{figure}

\begin{figure}
\epsscale{.90}
\plotone{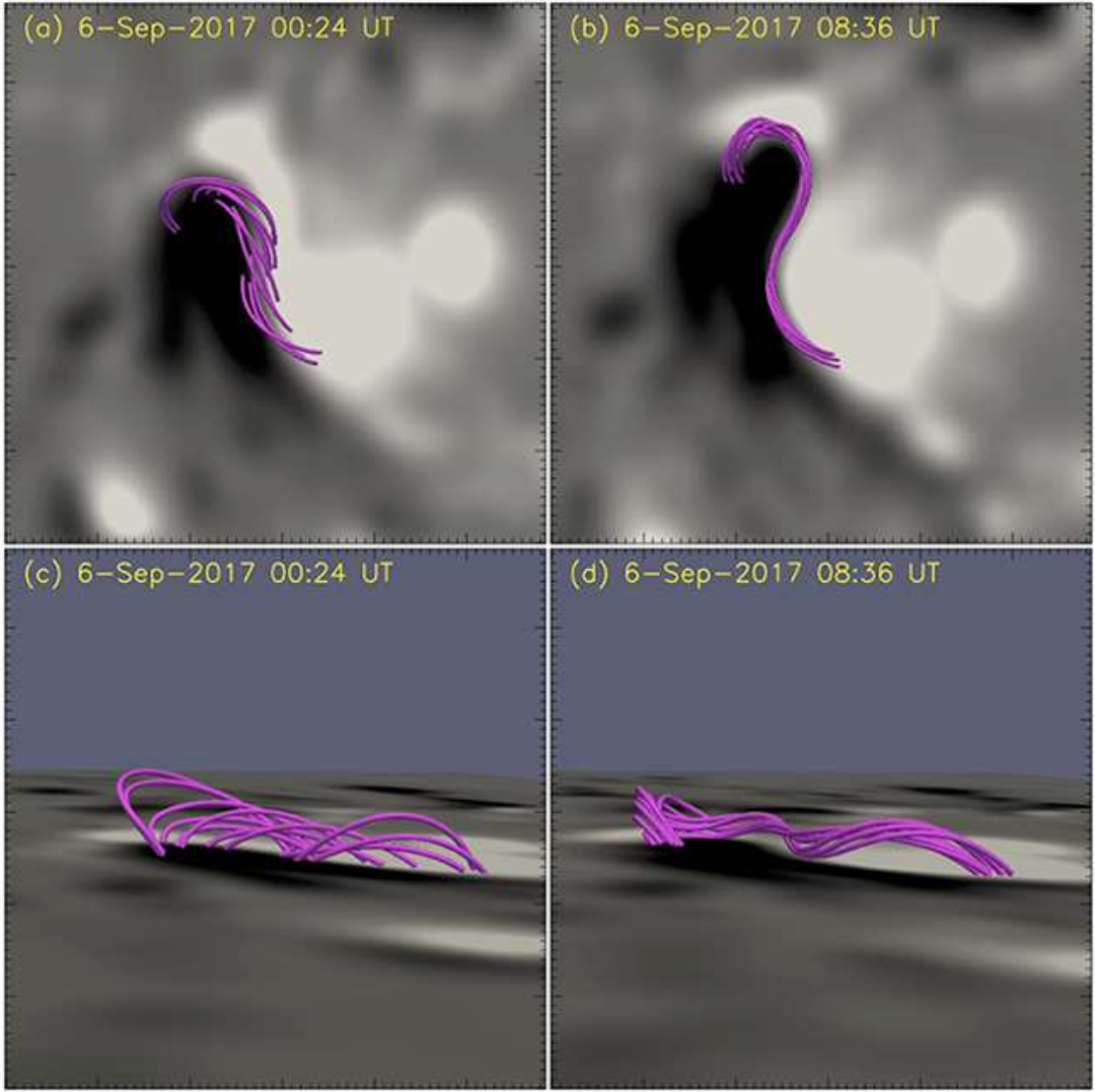}
\caption{ Extrapolation of the magnetic structure along the polarity inversion line between sunspots S1 and S2 superimposed on the longitudinal magnetic fields observed by SDO/HMI at 00:24 UT and 08:36 UT on 2017 September 6. Figs. 3(a) and 3(b) show the configuration of the first flux rope seen from top view. Figs. 3(c) and 3(d) show the configuration of the first flux rope seen from left side view. \label{fig1}}
\end{figure}

\begin{figure}
\centering
\includegraphics[angle=0,scale=0.8]{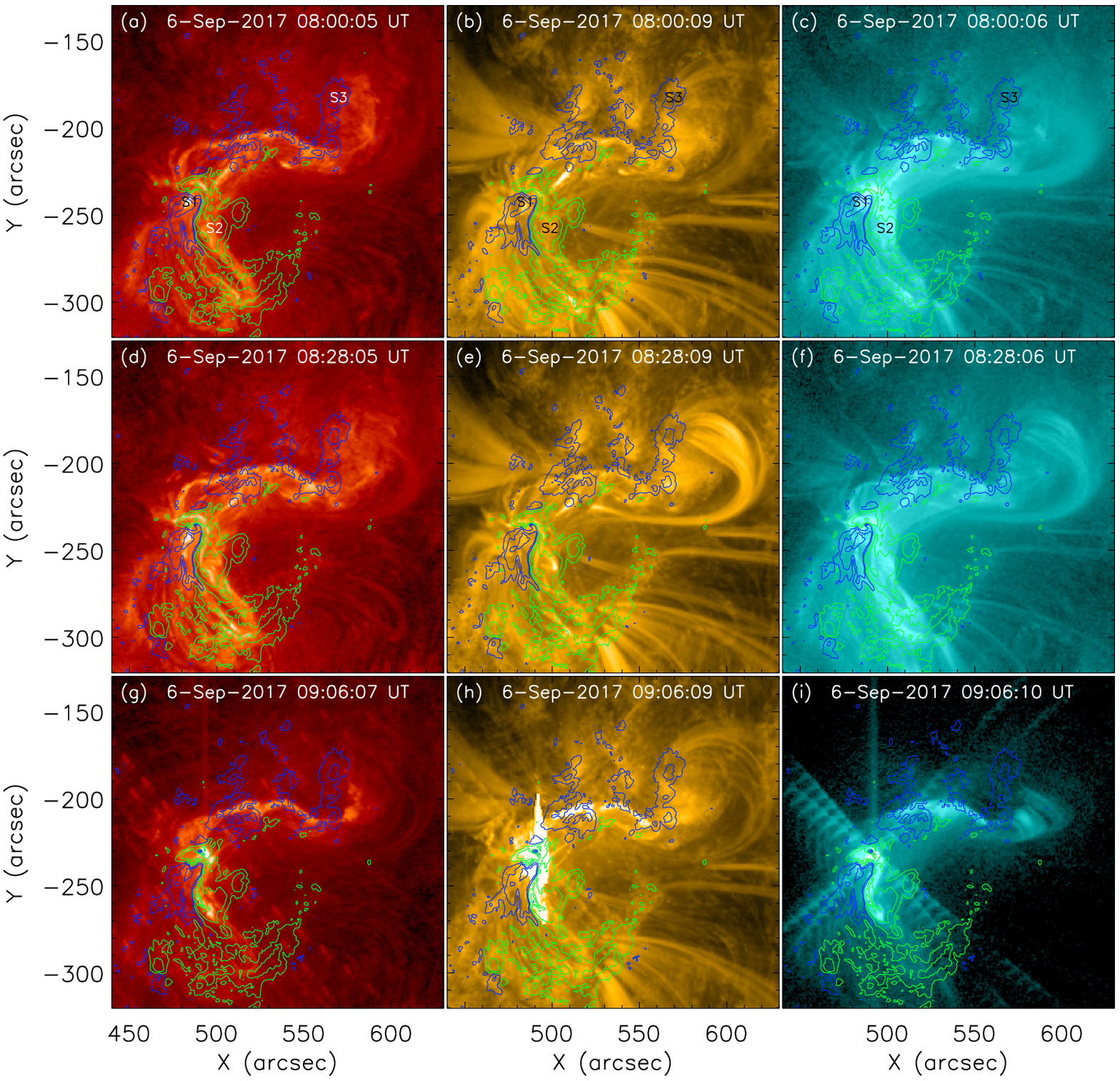}
\caption{Eruptive process of the X2.2 flare acquired at 304\AA, 171\AA, and 131\AA\ from the left column to the right one. All the images were overlaid the contours of the line of sight magnetograms. The green and blue contours indicate the positive and negative magnetic polarities. The contour levels of the magnetic fields are $\pm$350 G, $\pm$1000 G, and $\pm$1800 G.}
\end{figure}

\begin{figure}
\epsscale{.90}
\plotone{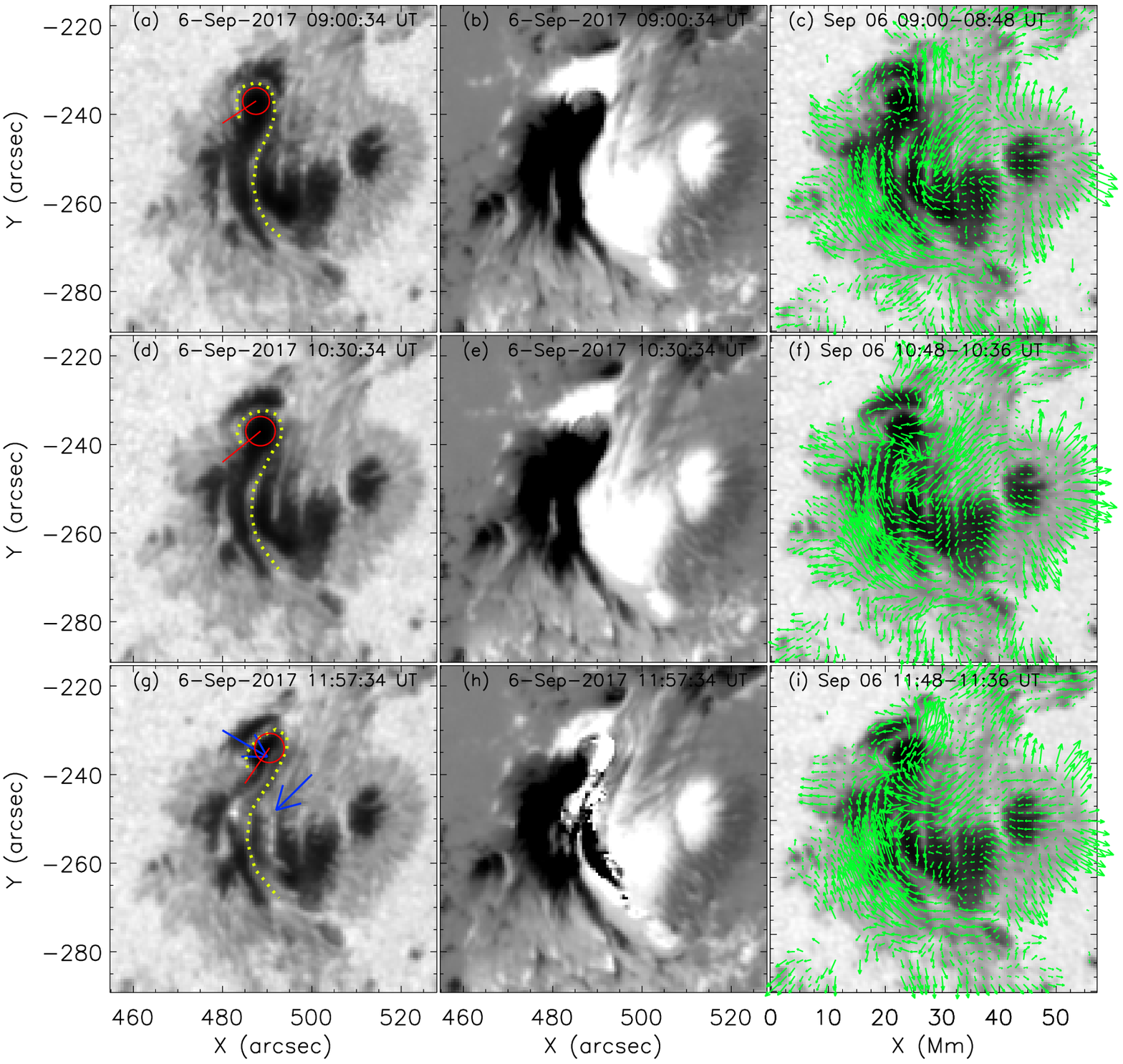}
\caption{Evolution of sunspots S1 and S2 in the continuum intensity images (the left column), the line of sight magnetograms (the middle column), and the velocity fields superimposed on the continuum intensity images (the right column). The yellow dotted lines indicate the change of the shape of the bright structure between sunspots S1 and S2. The green arrows in Figs. 4(c), 4(f), and 4(i) show the directions of the transverse flow fields. The positions of the enhancement of the intensity in the continuum intensity images were marked by two blue arrows at the maximum of the X9.3 flare in Fig. 4(g). The red circle covers the north part of sunspot S1 and is used to trace the rotation of the north part of the S-shaped structure with time. The red lines mark the position of the head of the S-shaped structure that rotated around the center of the north part of sunspot S1. \label{fig1}}
\end{figure}

\begin{figure}
\epsscale{.90}
\plotone{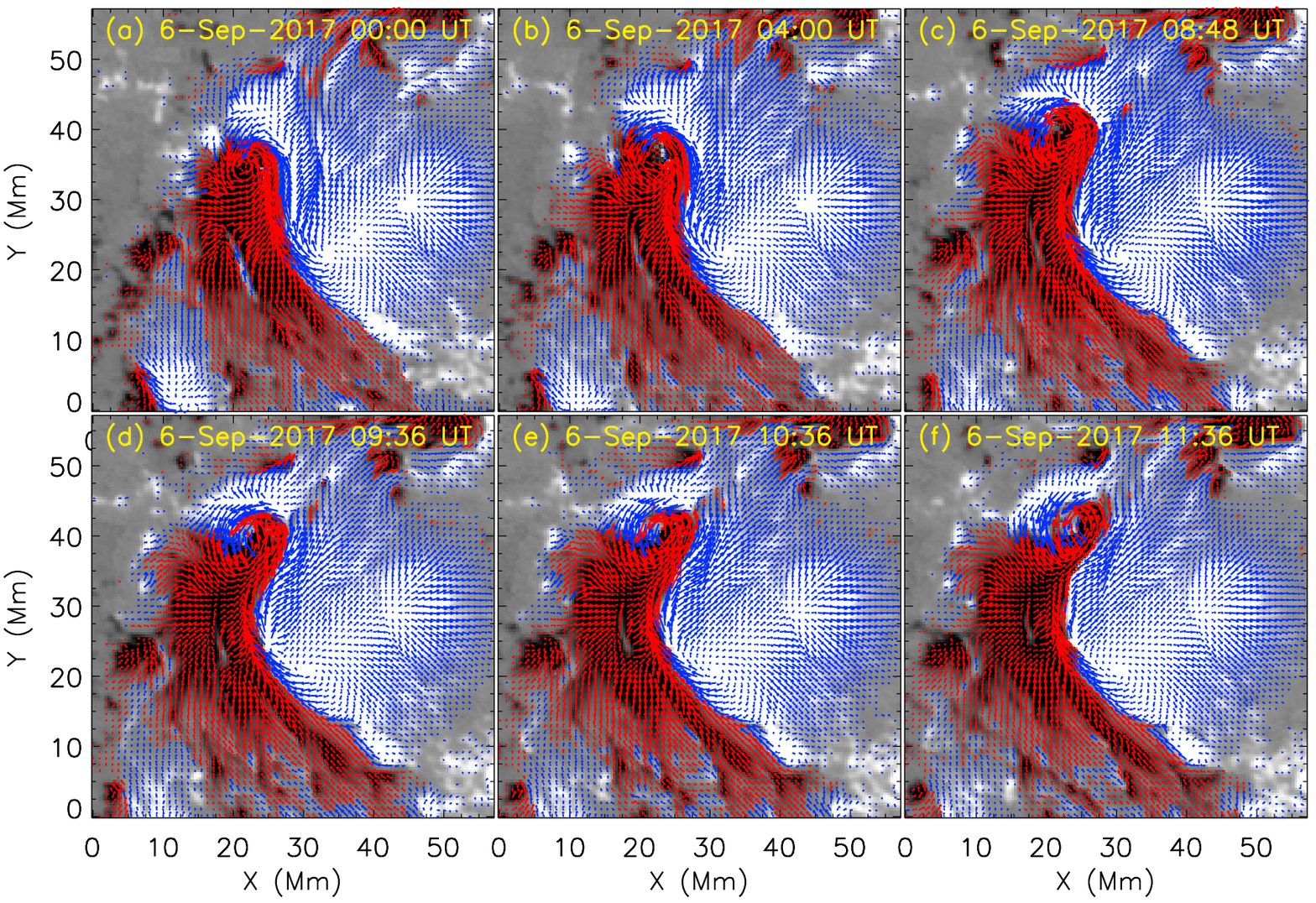}
\caption{Evolution of vector magnetograms of sunspots S2 and S3 observed by SDO/HMI from 00:00 UT to 11:36 UT on 2016 September 09. The red arrows show the transverse magnetic fields of negative polarity and the blue arrows show the transverse magnetic fields of positive polarity. \label{fig1}}
\end{figure}

\begin{figure}
\epsscale{.90}
\plotone{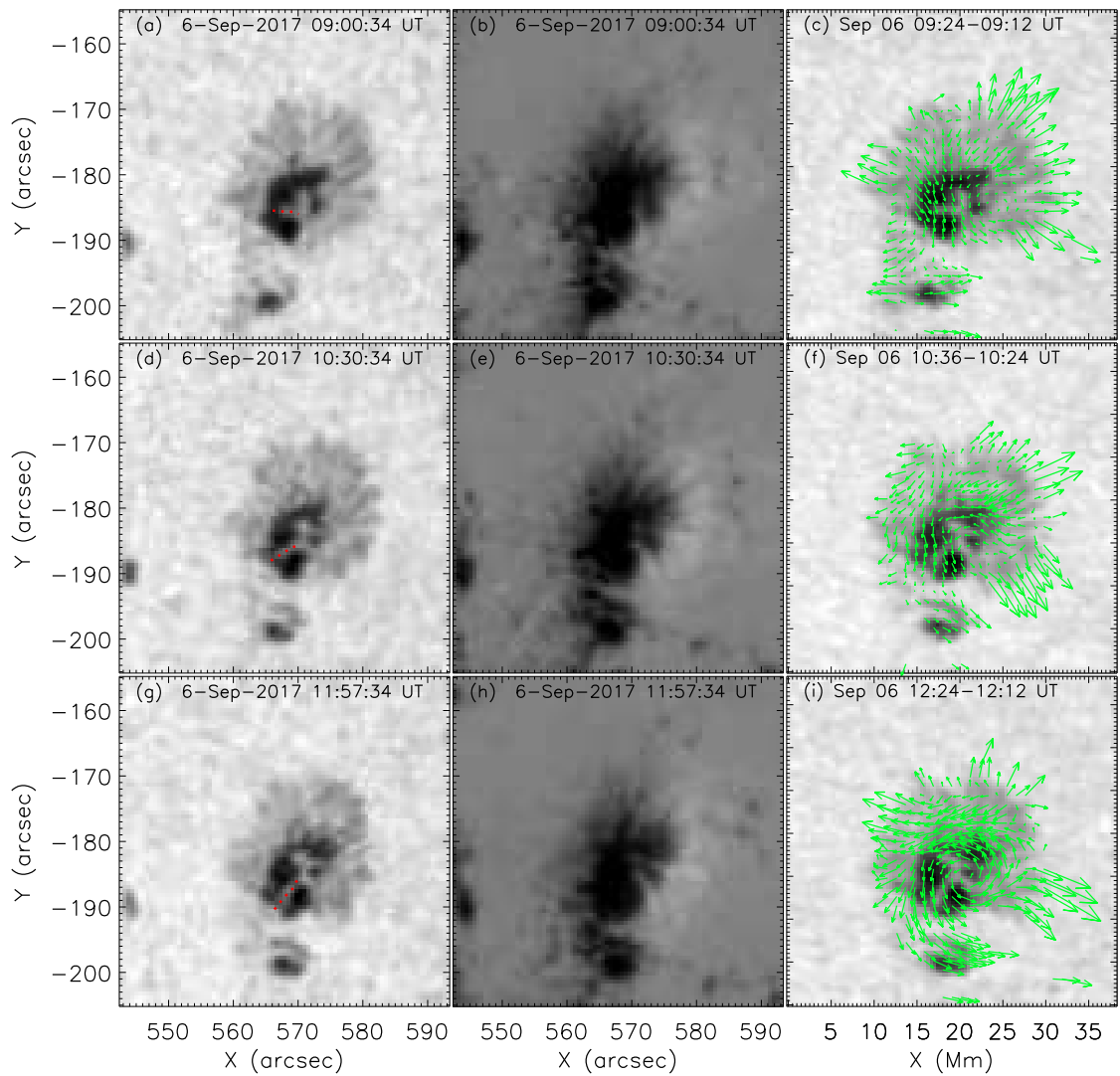}
\caption{Evolution of sunspot S3 in the continuum intensity images (the left column), the line of sight magnetograms (the middle column), and the velocity fields superimposed on the continuum intensity images (the right column). The red dotted lines indicate the change of the shape of the bright structure in the umbra. The green arrows in Figs. 5(c), 5(f), and 5(i) show the directions of the transverse flow fields. \label{fig1}}
\end{figure}

\begin{figure}
\epsscale{.90}
\plotone{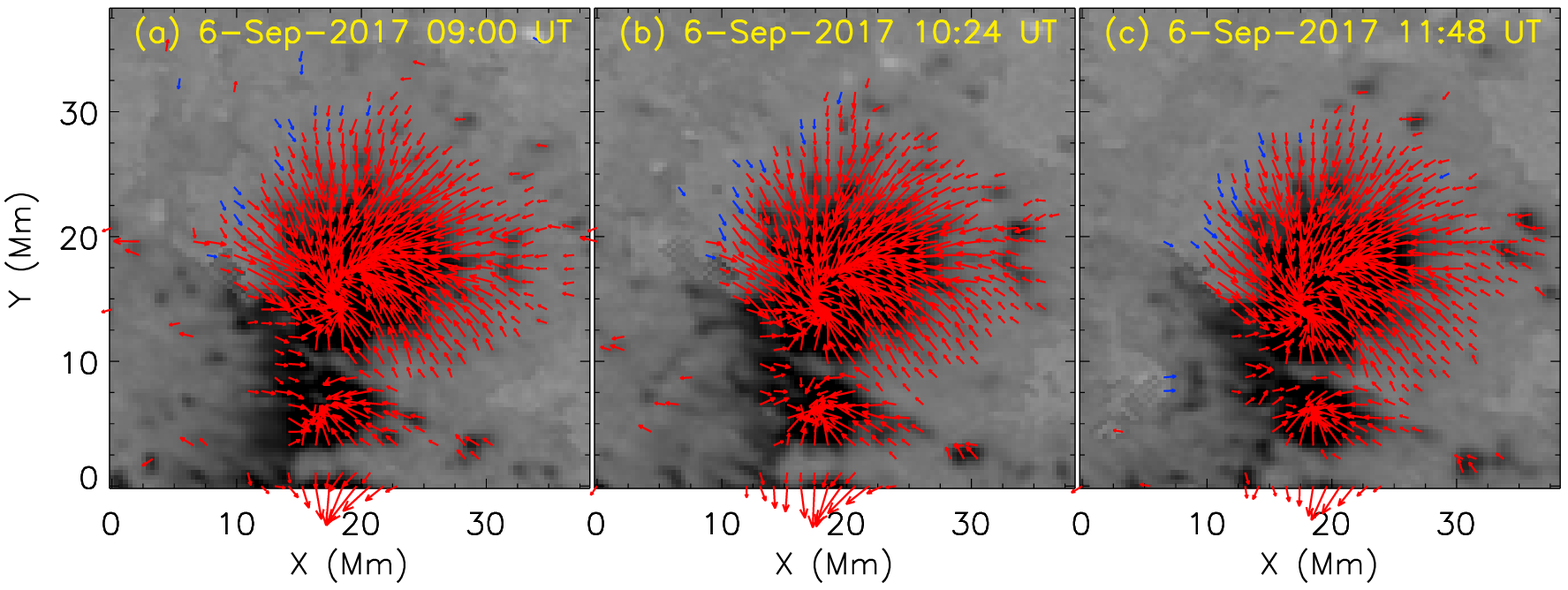}
\caption{Evolution of vector magnetograms of sunspot S3 observed by SDO/HMI from 09:00 UT to 11:48 UT on 2016 September 09. The red arrows show the transverse magnetic fields of negative polarity and the blue arrows show the transverse magnetic fields of positive polarity.  \label{fig1}}
\end{figure}

\begin{figure}
\epsscale{.90}
\plotone{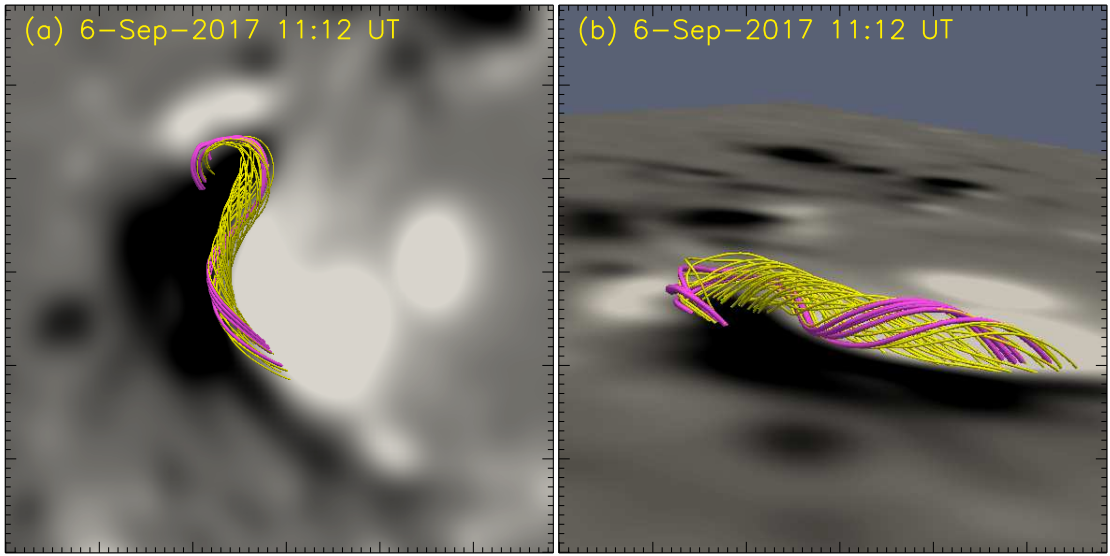}
\caption{Extrapolation of the magnetic structure along the polarity inversion line between sunspots S1 and S2 superimposed on the longitudinal magnetic fields observed by SDO/HMI at 11:12 UT on 2017 September 6. Figs. 9(a) shows the configuration of the first flux rope seen from top view. Figs. 9(b) show the configuration of the second flux rope seen from left side view.  \label{fig1}}
\end{figure}

\begin{figure}
\epsscale{1.0}
\plotone{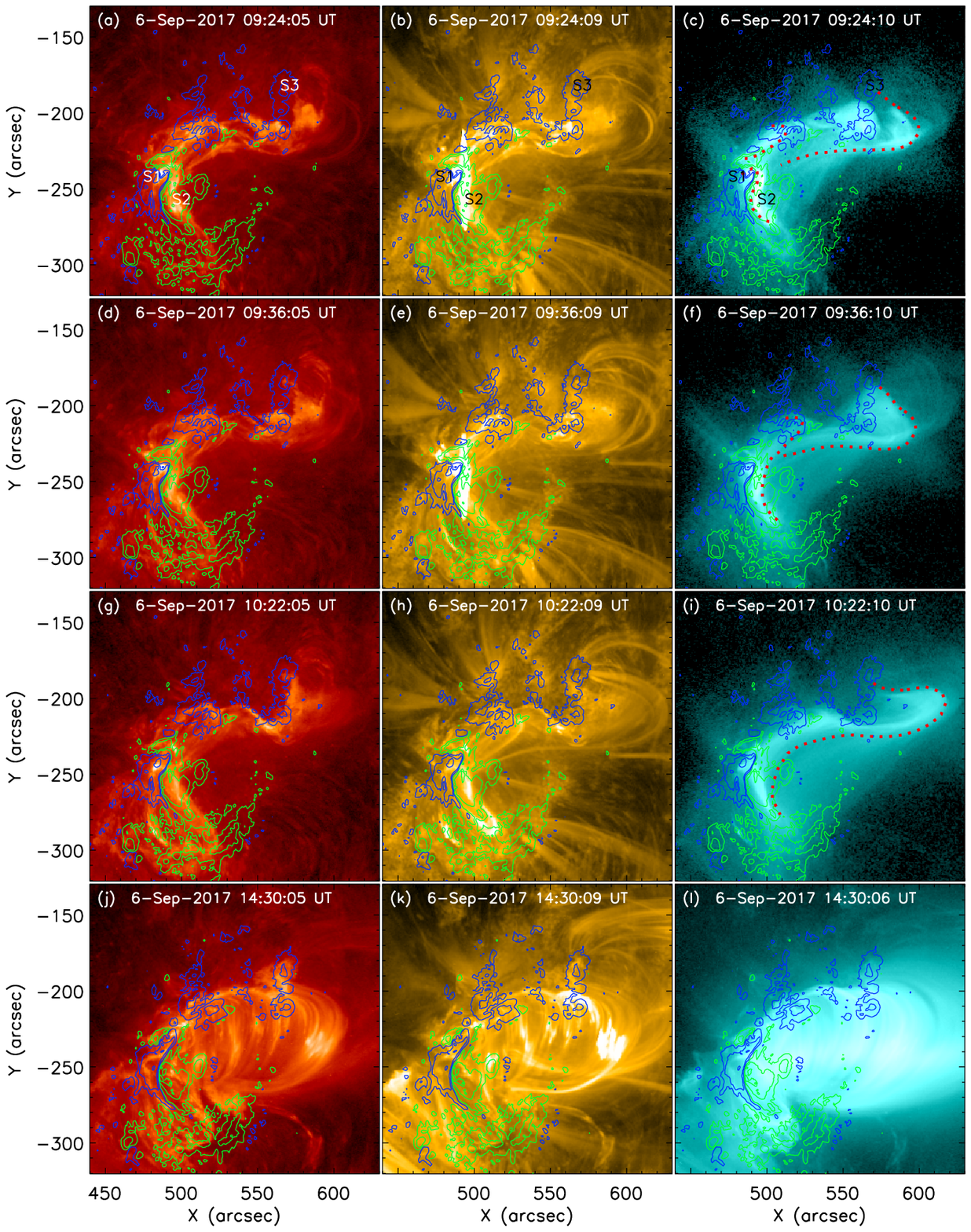}
\caption{Eruptive process of the X2.2 flare acquired at 304\AA, 171\AA, and 131\AA\ from the left column to the right one. All the images were overlaid the contours of the line of sight magnetograms. The green and blue contours indicate the positive and negative magnetic polarities. The contour levels of the magnetic fields are $\pm$350 G, $\pm$1000 G, and $\pm$1800 G. The three red dotted lines in Fig. 5(c) indicate the three parts of magnetic structures that were involved in the magnetic reconnection. \label{fig1}}
\end{figure}

\begin{figure}
\epsscale{.80}
\plotone{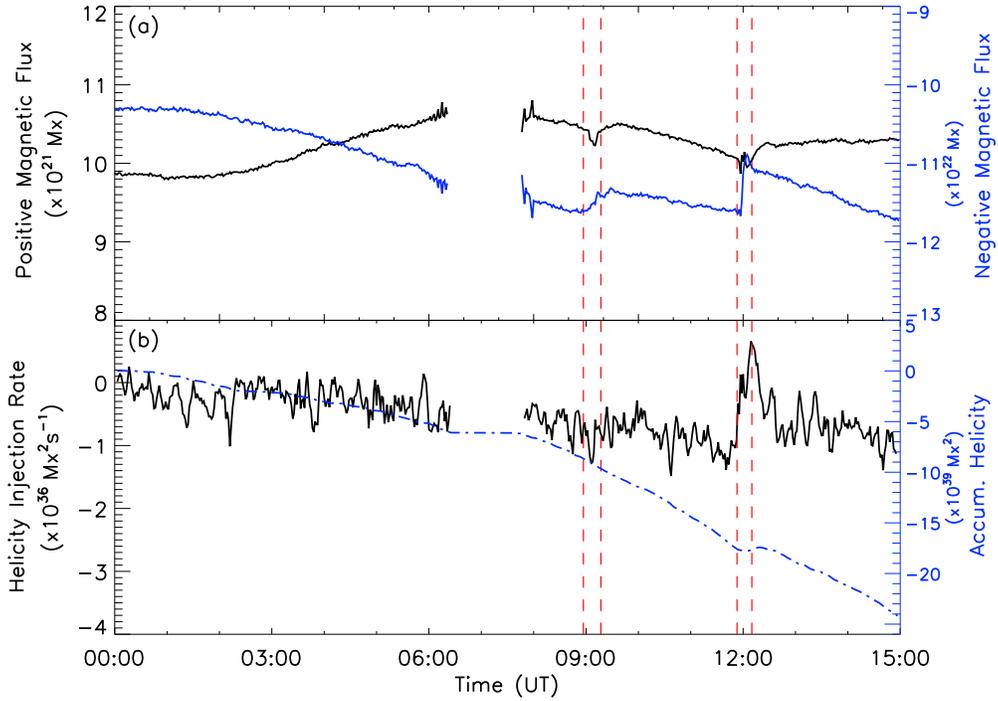}
\caption{Evolution of magnetic flux, the helicity injection rate, and the helicity accumulation in the whole active region. (a): The evolution of positive and negative magnetic flux. The black and blue lines indicate the change of the positive and negative magnetic flux on 2017 September 6 in Fig. 10(a). (b): The time profile of the helicity injection rate (black line in Fig. 10(b) and the helicity accumulation(blue line in Fig. 10(b)  in the whole active region. The two groups of the vertical dashed red lines indicate the starting and ending time of the two X-class flares.\label{fig1}}
\end{figure}

\begin{figure}
\epsscale{.60}
\plotone{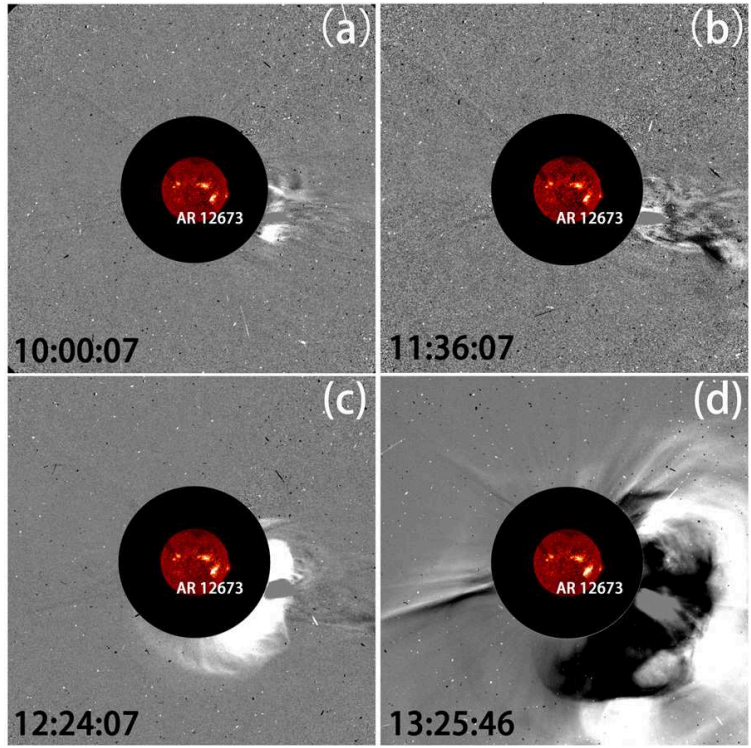}
\caption{Two CMEs associated with the two X-class flares observed by LASCO C2. (a) and (b): The first CME associated with the X2.2 flare. (c) and (d): The second CME associated with the X9.3 flare. The LASCO C2 images superimposed by the full disk 304 \AA\ images. The bright region in the south hemisphere in 304 \AA\ image was the location of the AR NOAA 12673.\label{fig1}}
\end{figure}
%\begin{figure}
%\epsscale{.80}
%\plotone{fig8.eps}
%\caption{ \label{fig1}}
%\end{figure}

\end{document}